\documentclass[%
 reprint,
 amsmath,amssymb,
 aps,
]{revtex4-2}

\usepackage{graphicx}
\usepackage{dcolumn}
\usepackage{bm}
\usepackage{epstopdf}
\usepackage[utf8]{inputenc}
\usepackage{xcolor}
\usepackage[export]{adjustbox}

\begin{document}

\preprint{APS/123-QED}

\title{Experimental implementation of laser cooling of trapped ions \\ in strongly inhomogeneous magnetic fields}

\author{Christian Mangeng}
\author{Yanning Yin}
\author{Richard Karl}
\author{Stefan Willitsch}
 \email{stefan.willitsch@unibas.ch}
\affiliation{%
 Department of Chemistry, University of Basel,  Klingelbergstrasse 80, 4056 Basel, Switzerland}%

\date{\today}

\begin{abstract}
We demonstrate the Doppler laser cooling of $^{40}$Ca$^+$ ions confined in a segmented linear Paul trap in the presence of a strong quadrupolar magnetic field generated by two permanent ring magnets. Magnetic field gradients of 800 to 1600~G/mm give rise to a highly position-dependent Zeeman shift on the energy levels of the trapped ions. Efficient laser cooling is demonstrated using two 397~nm cooling laser beams with appropriate wavelengths and polarizations and one 866~nm repumper laser beam. Coulomb crystals of ions are found to exhibit similar secular temperatures to those trapped in absence of the magnetic field. In addition, the position dependency of the Zeeman effect is used to generate a map of the field strength. This work forms the basis for developing hybrid trapping experiments for cold ions and neutral molecules that consist of an ion and a magnetic trap to study cold interactions between these species, and opens up new possibilities for quantum-science experiments that employ trapped ions in inhomogeneous magnetic fields.

\end{abstract}

\maketitle

\section{\label{section_introduction}Introduction}

\begin{figure*}[!t]
\includegraphics{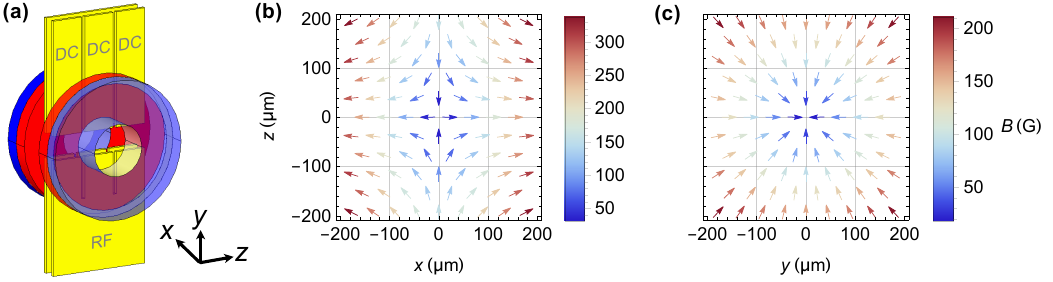}
\caption{\label{fig:trap}(a) Scheme of the segmented linear Paul trap enclosed between two permanent ring magnets generating a strong quadrupolar magnetic field. For clarity, the front magnet is rendered transparent. (b) Magnetic vector fields $\Vec{B}$ in the $xz$- and (c) $yz$-plane of the trapping region. The field gradients are 800 G/mm along the $y$ and $z$ directions and 1600 G/mm along the $x$ direction. The color code indicates the magnetic field strength $B$. The field vanishes at the center.}
\end{figure*}

Laser-cooled trapped ions are well-established platforms that find applications across diverse research fields such as quantum information and computation \cite{bruzewicz2019trapped, monroe2013scaling, haffner2008quantum, wineland2009quantum}, atomic clocks \cite{ludlow2015optical}, fundamental physics studies \cite{safronova2018search} and cold chemistry \cite{willitsch2012coulomb}. Although in Penning traps homogeneous magnetic ($B$) fields with typical strengths of several T are used to provide two-dimensional confinement of ions \cite{thompson2009applications}, adding magnetic fields to radiofrequency (RF, ``Paul" \cite{major05a}) traps, which confine ions only with electric fields, is less common because the resulting shifts and splittings of energy levels due to the Zeeman effect \cite{foot05a} considerably complicate laser cooling. So far, only the use of relatively weak homogeneous or inhomogeneous magnetic fields with modest field gradients has been reported for Paul traps. For example, homogeneous fields of strength 1–10 G are routinely implemented in Paul traps to define the quantization axis. Gradient magnetic fields have been employed in trapped-ion quantum-science experiments. Both static \cite{wang09b,johanning09a,khromova12a,weidt16a} and oscillating field gradients \cite{ospelkaus11a,warring13a,harty16a,hahn19a,zarantonello19a,srinivas21a} have been utilized to achieve laser-free entangling gates and enable the addressing of individual ions. Notably, these experiments employed relatively small gradients ranging from approximately 2.5 to 250~G/mm, and no significant impact on the efficiency of the laser cooling of the ions was reported. Theoretical studies on advanced cooling techniques within the Lamb-Dicke regime proposed incorporating magnetic field gradients \cite{albrecht2011enhancement}. Moreover, magnetically-assisted Sisyphus processes have been suggested for the cooling of molecular anions \cite{yzombard2015laser}. 

On the other hand, overlapping Paul traps with more strongly inhomogeneous magnetic fields, e.g. quadruplolar fields generated by permanents magnets, could open a door to more advanced applications in quantum-science experiments using trapped ions with highly position-dependent Zeeman-shifted energy levels, as well as in the study of cold ion-neutral collisions based on hybrid trapping systems which combine magnetic traps for neutral species and Paul traps for ions. Over the past years, hybrid traps of ions and neutrals \cite{willitsch2015ion, tomza2019cold} have emerged as versatile tools for studying interactions between these species at very low temperatures with much longer interaction times in comparison with experiments using molecular beams. Hybrid traps of ions and neutral atoms have been realized by overlapping Paul traps with magnetic \cite{zipkes10a}, optical-dipole \cite{schmid10a,meir16a,joger17a} or magneto-optical traps \cite{smith05a,grier09a,hall11a,rellergert11a, ravi11a, haze13a, jyothi19a} for atoms. However, realizing a hybrid trap of ions and neutral molecules by overlapping a Paul trap with a magnetic trap is less straightforward, since magnetic traps for molecules often exhibit much stronger inhomogeneous fields than those used for trapping atoms due to their higher temperatures \cite{hoekstra07a, tsikata10a, haas19a}, resulting in
highly position-dependent Zeeman splitting of ions that affects the ion motion and the efficiency of laser cooling of ions. 

Recently, we have studied the problem of laser cooling of ions confined in Paul traps in the presence of strongly inhomogeneous quadrupolar magnetic fields theoretically using molecular-dynamics simulations \cite{karl23}. There, the influence of the magnetic fields on the trapping and laser cooling of a single Ca$^+$ ion was explored, and schemes for laser cooling of single ions under ideal conditions (in which the center of the ion trap and the center of the magnetic field are perfectly overlapped) and configurations in which the two centers exhibit spatial displacements were suggested. 

Here, we demonstrate experimentally that assemblies of $^{40}$Ca$^+$ ions can be efficiently laser cooled in a segmented linear Paul trap exposed to a quadrupolar $B$ field with gradients exceeding 800 G/mm. The field is produced by a pair of permanent ring magnets. In addition, we make use of the fact that the Zeeman splittings become highly position-dependent in this environment to map out the magnetic field and to probe spatial mismatches between the center of the ion trap and the center of the magnetic field. The present setup is an essential prerequisite for realizing a hybrid trapping system in which cold collisions and reactions between ions and neutral molecules can be studied, which could also open new possibilities for trapped-ion experiments using magnetic fields with high field gradients.

\section{\label{section_experiment}Experiment}

\begin{figure}[!b]
\includegraphics{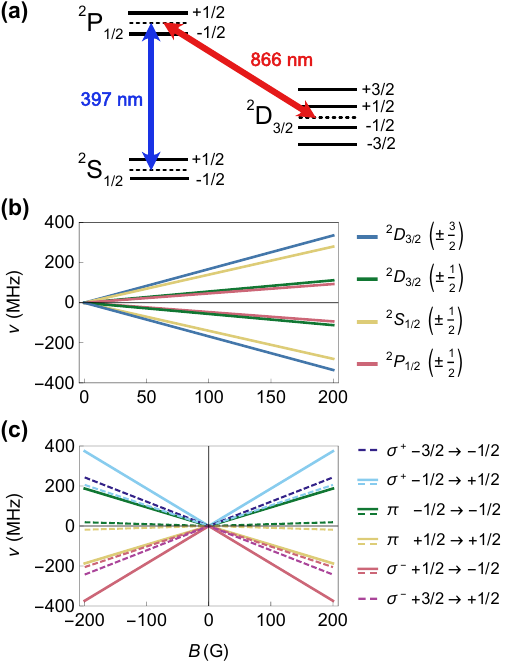}
\caption{\label{fig:zeeman_splits} 
(a) Schematic energy-level diagram of the $^{40}$Ca$^+$ cooling cycle with (solid lines) and without (dashed lines) Zeeman splitting. The levels are labeled by their magnetic quantum number $M$.
(b) Shifts of these levels (in units of frequency $\nu$) as a function of the magnetic field strength $B$. 
(c) Frequency shifts of the corresponding transitions in the magnetic field. Solid (dashed) lines correspond to Zeeman transitions within the $^2S_{1/2}\rightarrow {}^2P_{1/2}$ ($^2D_{3/2}\rightarrow {}^2P_{1/2}$) manifolds. The primary 397~nm cooling transition is split into four (two $\sigma$ and two $\pi$) and the 866~nm repumping transition into six (four $\sigma$ and two $\pi$) lines.}
\end{figure} 

The segmented linear-quadrupole RF trap used in the present study consisted of two RF plate electrodes and DC electrode segments as shown schematically in Figure \ref{fig:trap}(a). All electrodes were fabricated from titanium and covered with a 2 $\mu m$ gold layer. The DC electrode segments (width 4.5~mm, thickness 0.5~mm) were separated by 0.5~mm wide gaps along the $z$ direction (see Figure \ref{fig:trap}(a) for the coordinate system used throughout this study). The RF electrodes were not segmented but featured identical dimensions around the trap center to mimic the shape of the DC electrodes. The separations between the surfaces of the electrodes along the radial, i.e., $x$ and $y$ directions, were 1.5 and 3.5~mm, respectively, which resulted in a rectangular cross section of the trap in the $xy$-plane. For symmetry reasons, the RF fields vanish along the central longitudinal axis of the trap (the ``RF null line"). The trap was operated at an RF frequency of 10~MHz, RF amplitudes of 210~V to 300~V and static voltages of $\leq$10~V applied to the endcap electrodes.

Two permanent PrFeB ring magnets (Vakuumschmelze) with a remanence of 1.4~T (measured with a relative error of $\pm5\%$) were mounted on both sides of the ion trap at a distance of 0.5~mm from the outer surfaces of the electrodes. The magnets featured inner and outer diameters of 7.5 and 20~mm, respectively, and a thickness of 4.5~mm. Their central connecting axis was mechanically aligned to the center of the ion trap. The calculated magnetic vector fields generated by the magnets are presented in Figures \ref{fig:trap}(b,c) assuming perfect alignment and shapes of the magnets. The magnets form a quadrupolar field which vanishes in their common center (defined as the origin of the cartesian coordinate system in Figure \ref{fig:trap}). Two 30 $\mu$m thin grounded copper meshes with a line spacing of 250~$\mu$m were glued in between the magnets and the ion trap to provide electrical shielding. The meshes also served as a convenient ruler to estimate distances inside the trap during experiments with a relative accuracy of about 10\%.

Ions displaced from the magnetic center experience Zeeman splittings of their energy levels as shown in Figure \ref{fig:zeeman_splits}(a) for the levels system relevant for the Doppler laser cooling of Ca$^+$. The three-level system splits up into eight Zeeman components labeled by their magnetic quantum number $M$. Their energy shifts in the $B$-field are plotted for a field strength of up to 200 Gauss in Figure \ref{fig:zeeman_splits}(b). These field strengths correspond to ion displacements of up to 250~$\mu$m along the $y$/$z$ direction (field gradient 800 G/mm) and 125~$\mu$m along the $x$ direction (1600 G/mm). The number of possible transitions among these eight levels increases from two to ten, of which four derive from the $(4s)~^2S_{1/2}\rightarrow (4p)~^2P_{1/2}$ cooling transition at 397~nm (two $\sigma$ ($\Delta M=\pm1$) and two $\pi$ ($\Delta M=0$) components), and six from the $(3d)~^2D_{3/2}\rightarrow (4p)~^2P_{1/2}$ repumper transition at 866~nm (four $\sigma$ and two $\pi$ components), as shown in Figure \ref{fig:zeeman_splits}(c). In such a strongly inhomogeneous field, Doppler laser cooling with one 397~nm and one 866~nm laser becomes increasingly inefficient for ions displaced only a few tens of $\mu$m away from the zero of the magnetic field in the center between the magnets. According to \cite{karl23}, at least one additional 397~nm laser and a proper choice of wavelengths and polarizations of these lasers are required to cool the ions efficiently under such conditions.

\begin{figure}[!t]
\includegraphics[scale=0.45]{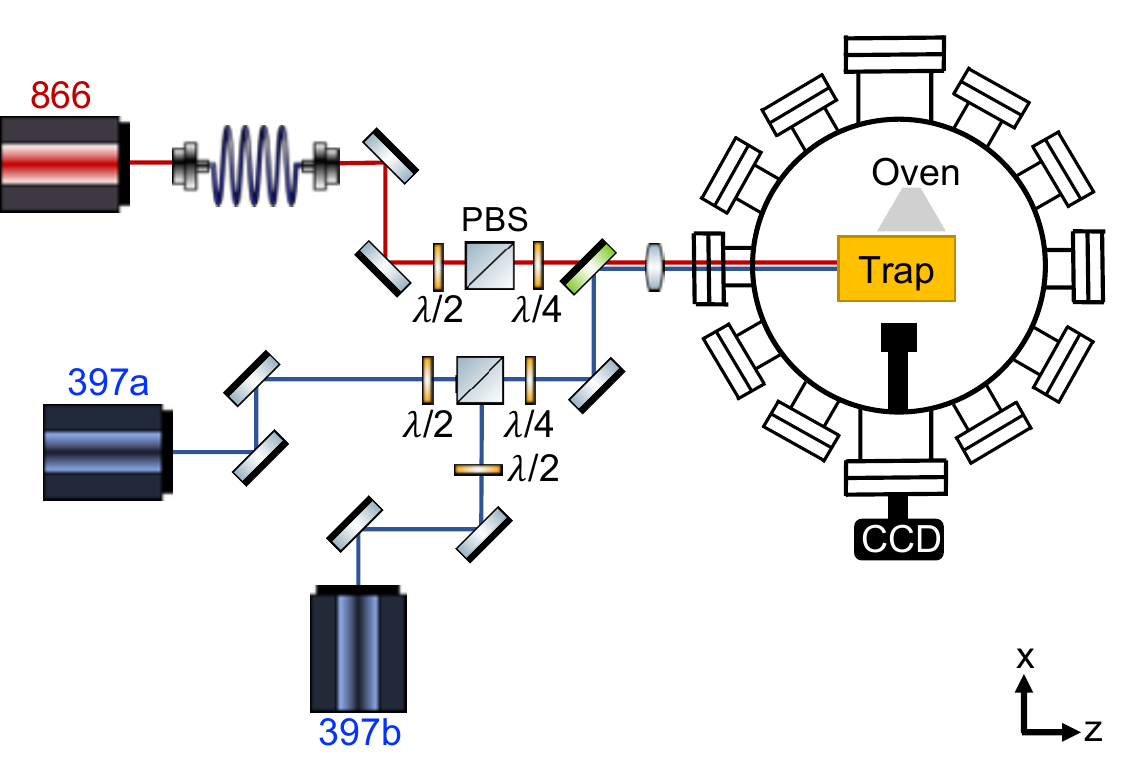}
\caption{\label{fig:experimental_setup} Scheme of the experimental setup used for laser cooling of trapped Ca$^+$ ions in a strongly inhomogeneous magnetic field. Details are given in the text.}
\end{figure}

The setup of the present experiment is shown in Figure \ref{fig:experimental_setup}. The ion trap is located at the center of a vacuum chamber at a pressure of $<5\times10^{-10}$~mbar. Ca atoms were evaporated from an oven and ionized inside the trap by laser beams at 375~nm and 423~nm (not shown in the figure) to generate Ca$^+$ ions. Two 397~nm cooling laser beams, labeled as 397a and 397b, as well as an 866~nm repumper laser beam were introduced into the trap along its longitudinal axis to cool the ions. Both 397~nm lasers beams were set to a power of 0.8~mW before entering the vacuum chamber and focused to a $1/{e^2}$ diameter of 500~$\mu$m at the position of the ions while the 866~nm laser was operated at 5.5 mW with a $1/{e^2}$ diameter of 750~$\mu$m. For each of these lasers, the polarization was freely tuneable using combinations of $\lambda$/2 plates, polarizing beamsplitters (PBS) and $\lambda$/4 plates as indicated in Figure \ref{fig:experimental_setup}. The beam 397a (397b) was transmitted (reflected) at a PBS, which lead to horizontal (vertical) linear polarization. With the $\lambda$/4 plate positioned after the PBS, the polarization could be tuned between linear, left-handed circular or right-handed circular to drive either $\pi$, $\sigma^-$ or $\sigma^+$ transitions. As a consequence of the shared PBS, 397a and 397b always possessed opposite linear or circular polarizations. In the trap, the laser-cooled ions assembled into ordered structures, referred to as Coulomb crystals \cite{willitsch2012coulomb}, as a consequence of the equilibrium between Coulomb repulsion and the confining electric fields in the trap. Their spatially resolved laser-induced fluorescence was collected by a lens system and recorded by a charge-coupled device (CCD) camera.

\section{\label{section_results}Results and Discussion}

\subsection{Mapping the magnetic field using the Zeeman effect in laser-cooled Ca$^+$}
\label{sec:cooling}

\begin{figure*}[t]
\includegraphics{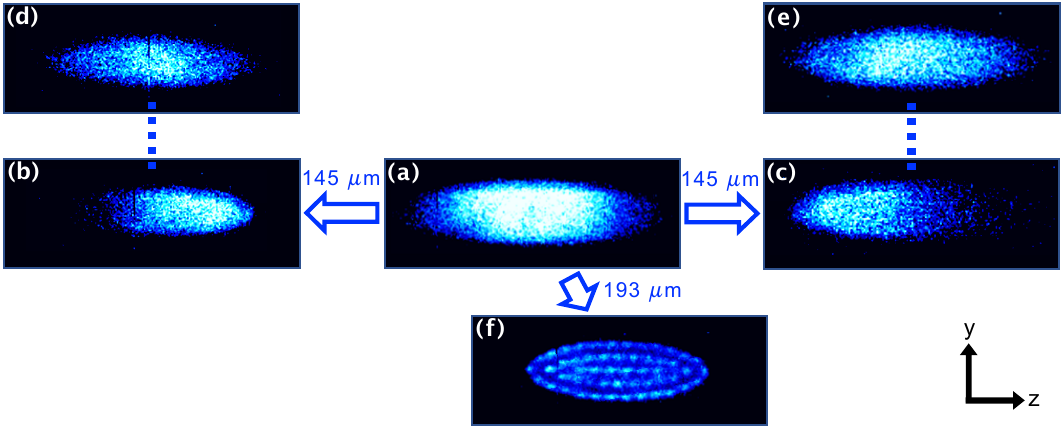}
\caption{\label{fig:field_map} False-color resonance-fluorescence images of Coulomb crystals in the RF trap within the quadrupolar magnetic field generated by two ring magnets illustrating the position dependency of the Zeeman shifts. (a) Crystal in the center of the magnetic trap cooled by only a single 397~nm primary cooling laser. (b) and (c) crystals moved along the $z$-direction  by $\pm145~\mu$m at constant cooling laser wavelength. (d) and (e) Translated crystals cooled by two 397~nm lasers with optimized wavelengths and polarizations. (f) Crystal in the ion trap center, located 193~$\mu$m away from the magnetic center, cooled by two lasers under appropriately optimized conditions. See text for details.}
\end{figure*}

The fixed position and magnetization of the ring magnets result in a static magnetic field that splits the energy levels of the ions based on their location in the trap. Conversely, every ion position has an attributed range of laser wavelengths at which the respective ion is cooled efficiently, reflected by the intensity of the resonance fluorescence of the ion at the specific location. As a consequence, one can use an ensemble of trapped ions to generate a spatial map of the magnetic field. With the present large field gradients, contrasts in the fluorescence signal are well resolvable with crystals containing about 100 ions.

\begin{figure}[b]
\includegraphics{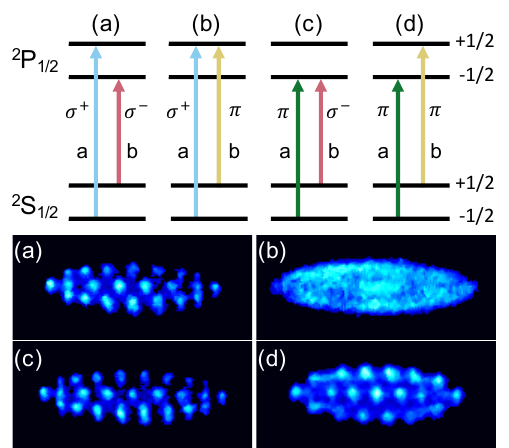}
\caption{\label{fig:transitions} False-color images of small Coulomb crystals obtained for driving different pairs of Zeeman components of the $^2P_{1/2}~\leftarrow~{}^2S_{1/2}$ cooling transition. Labels a and b denote the two lasers 397a and 397b to drive the transitions (see Figure \ref{fig:experimental_setup}). The color code for the transitions is the same as in Figure \ref{fig:zeeman_splits}(c).}
\end{figure}

We moved the ions within the trap by changing the DC and RF voltages and scanning the wavelengths of the cooling lasers. We located a position in the trap where stable crystals could be obtained using a single cooling laser with arbitrary polarization. We identify this location as the center of the magnetic trap, at which the field is close to zero and thus negligible Zeeman shifts occur. If the center of the crystal was aligned with the zero of the magnetic field, the region of maximum ion fluorescence was observed in the center of the crystal, as shown in Figure~\ref{fig:field_map}(a). The fluorescence intensity decayed symmetrically towards the edges of the crystal due to the increasing field-induced splitting of the cooling transition from its central frequency (Figure \ref{fig:zeeman_splits}(c)). 

Starting from the center of the magnetic trap, we moved the ions along the negative and positive $z$ axis by $\approx145~\mu$m as shown in Figs. \ref{fig:field_map}(b) and (c), respectively. At these positions, only a part of the ions in the crystal still showed appreciable resonance-fluorescence yields, while other parts of the crystal became increasingly dark reflecting the varying Zeeman shifts across the crystal within the gradient of the quadrupole magnetic field and the splitting of the cooling transition into several Zeeman components. Consequently, a single 397~nm laser was not sufficient anymore to efficiently cool the entire crystal, and additional laser beams were needed to address additional Zeeman cooling transitions. If the wavelengths and polarizations of the two cooling lasers were optimized carefully, the fluorescence was recovered across the entire crystal in either position (Figure \ref{fig:field_map}(d,e)). In this scenario, we found that only one of the four possible pairs of Zeeman cooling transitions indicated in Figure \ref{fig:transitions}(a) can be used to obtain a stable crystal: the combination of $\sigma^+$ and $\sigma^-$. This is because $\pi$ transitions are not allowed if the laser propagation axis aligns with the direction of the magnetic field as is the case for the present translation along the $z$ axis (Figure \ref{fig:trap}(b,c)). At the crystal positions indicated in Figs. \ref{fig:field_map}(b,c), we drove the $\sigma^+/\sigma^-$ transitions with right-(left-)handed circularly polarized cooling lasers 397a (397b) at a frequency shift of +276 ($-$276)~MHz from the center wavelength. 

Generally, we found that crystals located near the zero of the magnetic field showed stronger fluorescence compared to other locations, but appeared relatively hot so that individual ions were not resolved in the images. We attribute this to RF-induced heating at this position in the trap, suggesting that the zero positions of the magnetic and the RF fields do not perfectly overlap due to slight mechanical misalignments of the assembly. 

The coldest crystals for which positions of single particles could be resolved in the images were obtained by translating the ions from the magnetic center 180~$\mu$m along the negative $y$-direction and 70 $\mu$m along the positive $z$ axis, i.e. a diagonal distance of 193~$\mu$m. Along the $x$ direction, no significant adjustments were required. We conclude that this position corresponds to the ion trap center. The low temperature of the ions in the crystals trapped at this location is a consequence of minimised RF heating, which becomes larger if the distance from the center is increased. In the trap center, the ions could be cooled not only by driving $\sigma$, but also by $\pi$ transitions as the magnetic-field vectors now lie at an angle to the laser propagation direction. We found that cold crystals could be obtained by driving three out of the possible four combinations of Zeeman transitions as shown in Figure \ref{fig:transitions}. The $\sigma^+/\sigma^-$ pair of transitions was driven at frequency shifts of +333/$-$352~MHz and the $\pi$ pair with horizontal/vertical linearly polarized components at +162/$-$162~MHz. The $\pi/\sigma^-$ pair represented a combination of the two, that is, linear and left-handed circularly polarized components at +162/$-$352~MHz. Using the $\sigma^+/\pi$ pair of transitions, on the other hand, did not lead to a similarly cold crystal for unclear reasons (Figure \ref{fig:transitions}(b)).

We conclude from these results that the central axis joining the magnets is shifted with respect to the ion-trap center by 193~$\mu$m. However, the alignment of the magnets with respect to their common axis was precise enough to directly compare the experimentally determined Zeeman splittings with the calculated ones in Figure \ref{fig:zeeman_splits}(b,c).

For the crystals in Figure \ref{fig:field_map}(d,e), which were obtained at ($x$, $y$, $z$) = (0, 0, $\pm$145~$\mu$m), we calculate a theoretical field strength of 116~G according to Figs. \ref{fig:trap}(b,c). Consequently, for the $\sigma^+/\sigma^-$ pair of transitions used for cooling, theoretical frequency shifts of +217/$-$217~MHz are expected at this position. These predictions compare to the experimentally determined values of +276/$-$276~MHz. For the crystal in Figure \ref{fig:field_map}(f) obtained at ($x$, $y$, $z$) = (0, $-$180~$\mu$m, +70~$\mu$m), we calculated a field of 200~G, resulting in shifts of +374/$-$374 (measured: +333/$-$352)~MHz for $\sigma^+/\sigma^-$ and +187/$-$187 (+162/$-$162)~MHz for the $\pi$ pair. The differences between measured and theoretical shifts lie within the power-broadened linewidth of the 397~nm lasers which is calculated to be 95~MHz.

\subsection{Comparison of Coulomb crystals with and without the quadrupolar magnetic field}

\def\crystw{0.232}

\begin{figure}[!t]
\centering
\includegraphics{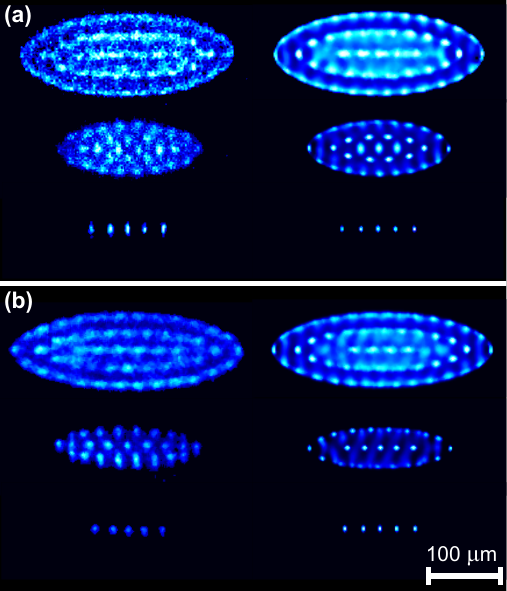}
\caption{\label{fig:crystals} Examples of false-color images of Ca$^+$ Coulomb crystals trapped (a) in absence and (b) in presence of a quadrupolar magnetic field near the ion trap center. Left: experimental crystal images, right: simulations. A medium, a small crystal and an ion string are shown in each case. Ion numbers and secular temperatures from top to bottom are for (a) 160, 70, 5 ions and 9, 7, $<$7 mK and for (b) 170, 50 and 5 ions and 9, 5, $<$5 mK. The images were averaged over 3 and 10 exposures for (a) and (b), respectively.}
\end{figure}

To assess the effect of the magnetic field on the cooling efficiency and final temperature of the ions, control experiments were performed without the magnets. A selection of images of crystals of different sizes obtained under these conditions, as well as their molecular-dynamics (MD) simulations, are shown in Figure \ref{fig:crystals}(a). The MD simulations were based on the methodology described in \cite{rouse15a}. Comparison between experimental and simulated crystal images allowed the extraction of ion numbers and secular temperatures \cite{willitsch2012coulomb}. 

We generated crystals with ion numbers between 100 and 200, which we define as medium size, smaller ones with less than 100 ions, as well as ion strings, for comparison in absence and presence of the field, as presented in Figure \ref{fig:crystals}(b).

In both cases, the crystals (and strings) exhibit similar secular temperatures at comparable ion numbers. This demonstrates that under the present conditions, two 397~nm lasers at appropriate frequencies are sufficient to cool crystals of the selected sizes efficiently in the presence of the magnetic field. It also confirms that only one 866~nm laser beam is needed to cover the $^2P_{1/2}~\leftarrow~{}^2D_{3/2}$ repumping transitions despite their large Zeeman splittings (estimated to span a total of $\pm$336~MHz ($^2D_{3/2}$, $M=\pm3/2$) at the ion trap center). The high intensity of the 866~nm laser in the present experiments sufficiently power broadened the transitions to be able to drive all relevant Zeeman components (Figure \ref{fig:transitions}). No dark Zeeman sublevels were observed in the optical cycling, consistent with the simulations \cite{karl23}.

\subsection{Motional frequencies of ions in the combined trap}

\begin{figure}[t]
\includegraphics[width=0.48\textwidth]{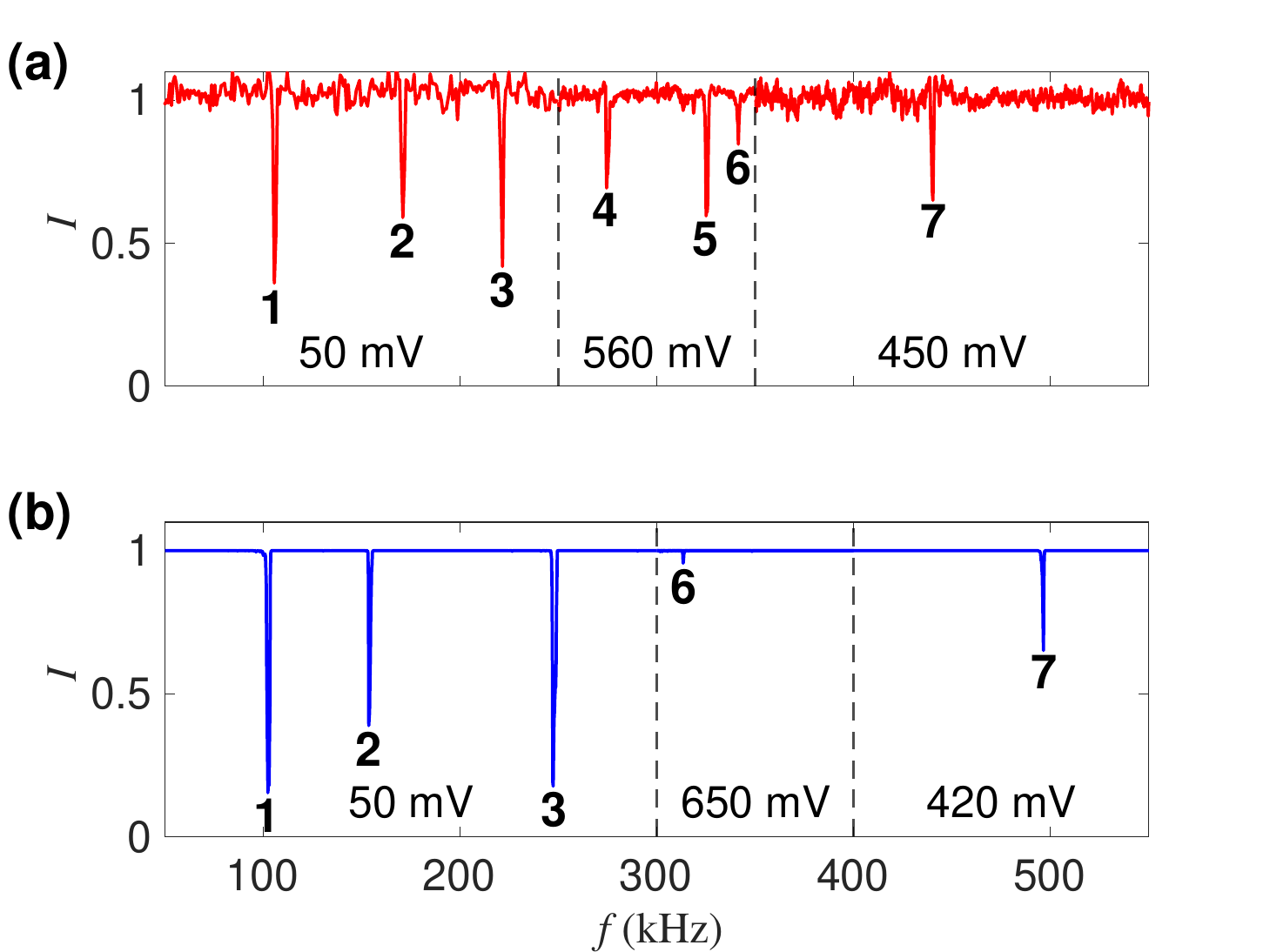}
\caption{\label{fig:resonance_spectra} Resonant-motional-excitation spectra showing the normalized fluorescence intensity $I$ of the ions as a function of the drive frequency $f$ for a medium-sized crystal in (a) the ion trap center and (b) the magnetic-field center. The peaks correspond to excitations of secular motions: (1) axial, (2, 3) radial, (4, 5) coupling between axial and radial, (6, 7) radial, second harmonic. Tickling voltages varied in different regions as indicated in order to probe all resonances.}
\end{figure}

The present setup enabled the examination of further location-dependent dynamics of the trapped ions. One example is the coupling of ion motions by the magnetic field, which we characterized by resonant excitation of the collective motion of the crystallized ions by the application of additional RF fields to the trap electrodes \cite{drewsen04a}. Motional resonances manifest themselves by a pronounced decrease of the resonance fluorescence if the frequency of the ion motion matches the frequency of the RF drive fields. Resonant-excitation spectra recorded at both the magnetic- and ion-trap centers are shown in Figure \ref{fig:resonance_spectra}. As the ions can move in three spatial directions, one axial and two radial frequencies were measured in both cases (peaks 1 to 3). In addition, the radial motions of the ions exhibited measurable second harmonics, which appear at double the frequencies of these motions (peaks 6 and 7). Features corresponding to the sum frequencies of the axial and radial motions were detected only for ions located in the ion trap center (peaks 4 and 5 in Figure \ref{fig:resonance_spectra}(a)), but are absent from the spectrum taken at the magnetic center (Figure \ref{fig:resonance_spectra}(b)). This suggests that the coupling of the axial and radial motions is predominantly induced by the magnetic field, as predicted in Ref. \cite{karl23}, rather than other mechanisms like Coulomb interactions or trap anharmonicities. The lower signal-to-noise ratio of the spectrum in the ion-trap center is due to the smaller resonance-fluorescence yield as only a subset of the possible cooling transitions is addressed (see Sec. \ref{sec:cooling}).

\section{\label{section_conclusion}Conclusion and Outlook}

The present study demonstrated efficient laser cooling of trapped $^{40}$Ca$^+$ ions in the presence of a strong quadrupolar magnetic field with gradients of 800 to 1600~G/mm. The ions reached similar temperatures at comparable ion numbers with and without magnetic field by using only one additional cooling laser to address specific Zeeman components of the cooling transition. We showed that the position-dependent Zeeman shifts can be used to map the magnetic field by identifying the wavelength difference of the cooling lasers and the corresponding driven transitions revealing a slight spatial mismatch between the centers of the ion and magnetic traps in our setup. "Hot" crystals could be obtained in the center of the magnetic trap by using only one cooling laser. Cold crystals at a lower resonance-fluorescence yield were obtained in the ion-trap center where the RF-induced heating is minimal. Magnetic-field-induced coupling between the axial and radial ion motions were observed under these conditions.

This work shows that a hybrid trapping system consisting of a Paul trap and a magnetic trap is experimentally feasible. Such a setup could be used to study cold collisions and reactions between ions and neutral molecules. Since magnetic field gradients are also widely used in trapped-ion quantum-science experiments, this work could open new possibilities for applications that benefit from the trapped ions with highly position-dependent Zeeman shifts.

\begin{acknowledgments}
We acknowledge financial support from the Swiss National Science Foundation (grant numbers 200020\_175533 and TMAG-2\_209193) as well as the University of Basel. Yanning Yin acknowledges support from the Research Fund of the University of Basel for Excellent Junior Researchers (grant nr. 3CH1051).

Christian Mangeng and Yanning Yin contributed equally to this work.
\end{acknowledgments}
\bibliographystyle{apsrev4-2}
\bibliography{_refs}

\end{document}